\documentclass[11pt]{article}
\usepackage{amsfonts}
 \usepackage{amsmath}

\usepackage{mathrsfs}
\usepackage{fancyhdr}
\usepackage{cite}
\usepackage{color}

\newcommand{\bgreen}{\begin{color}{green}}
\newcommand{\bred}{\begin{color}{red}}
\newcommand{\bblu}{\begin{color}{blue}}
\newcommand{\ecl}{\end{color}}

\usepackage[numbers,sort&compress]{natbib}
\begin{document}
\renewcommand{\theequation}{\thesection.\arabic{equation}}

\begin{center}
\LARGE\bf Algebro-geometric integration of the Q1 lattice equation via nonlinear integrable symplectic maps
\end{center}

\begin{center}
 Xiaoxue Xu$^{\rm 1}$, Cewen Cao$^{\rm 1}$, Frank W Nijhoff$^{\rm 2}$
\end{center}

\begin{center}
${}^{\rm 1}$ School of Mathematics and Statistics, Zhengzhou University, Zhengzhou, 450001, PR China\\
${}^{\rm 2}$ Department of Applied Mathematics, University of Leeds, Leeds LS2 9JT, UK\\
E-mail: frank@maths.leeds.ac.uk
\end{center}

\begin{abstract}
The Q1 lattice equation, a member in the Adler-Bobenko-Suris list of 3D consistent lattices, is investigated. By using the multidimensional consistency, a novel Lax pair for Q1 equation is given, which can be nonlinearised to produce integrable symplectic maps. Consequently, a Riemann theta function expression for the discrete potential is derived with the help of the Baker-Akhiezer functions. This expression leads to the algebro-geometric integration of the Q1 lattice equation, based on the commutativity of discrete phase flows generated from the iteration of integrable symplectic maps.
\end{abstract}


\section{Introduction }

In this paper we construct algebro-geometric solutions of the Q1 quadrilateral lattice equation,
\begin{equation}\label{eq:1.1}
\Xi^{(0,2)}\equiv \beta_{1}^{2}(\tilde{\bar{u}}-\tilde{u})(\bar{u}-u)-\beta_{2}^{2}(\bar{\tilde{u}}-\bar{u})(\tilde{u}-u)
+\delta^{2}\beta_{1}^{2}\beta_{2}^{2}(\beta_{1}^{2}-\beta_{2}^{2})=0,
\end{equation}
denoted by the symbol $\Xi^{(j,k)}$ where $j,k$ corresponds to the number of continuous and discrete variables respectively. In \eqref{eq:1.1}
we have adopted our preferred short-hand notation for lattice systems, i.e., partial difference equations for functions $u=u_{m,n}$
depending on two discrete independent variables $m,n\in\mathbb Z$, forming a regular lattice with coordinates
$(m,n)\in\mathbb Z^2$, and where elementary lattice shifts are denoted by $\tilde{u}=u_{m+1,n}, \bar{u}=u_{m,n+1}$.
Equation (\ref{eq:1.1}) is a member in the well-known ABS (Adler-Bobenko-Suris) list of 3D consistent lattices \cite{Adler}, where $\beta_1,\beta_2$ are (lattice)
parameters associated with the two lattice directions respectively, while $\delta$ is a fixed parameter. For $\delta\neq 0$ this equation
first appeared in the classification of \cite{Adler}, but the case $\delta=0$ (which we will denote by $({\rm Q}1)_0$) first appeared
in \cite{NC} where, due to the appearance of the canonical cross ratio of four variables, it was identified as a lattice version of the Schwarzian
Korteweg-de Vries (KdV) equation. In the previous papers, \cite{XuJiaNij,XuCZ}, we constructed algebro-geometric solutions of the $({\rm Q}1)_0$ equation, using the method
of symplectic maps arising from a nonlinearisation approach \cite{Cao1990,Cao1991}.
The present paper considers the $\delta$-parameter extension of that equation, which amounts to a significant departure from the $\delta=0$ case,
since in a sense it `lifts' the equation away from the KdV related lattice equations, cf. e.g. \cite{Hietarinta}, and
towards what one could call the Krichever-Novikov (KN) class, on which we say a bit more below. As a consequence of the presence of this parameter
the corresponding Lax pair is
sigificantly more complicated than the KdV type lattice equations, and various other simplifying features disappear. Hence, also the nonlinearisation
approach to the algebro-geometric solutions is significantly different from the one for the KdV class systems. With regard to explicit solutions of \eqref{eq:1.1}
relatively little is known so far: soliton solutions were constructed in \cite{NijAtkHiet}, while elliptic type solutions were presented in
\cite{NijAtk}, along with those of all members of the ABS list apart from the top equation Q4. Elliptic $N$-soliton type solutions of the latter equation
(where the equation itself is defined over an elliptic curve) were constructed in \cite{AtkNij}. So far, to our knwoledge, no explicit results exist
about the algebro-geometric solutions of any of the $\delta$-parameter equations in the ABS list for higher genus $g>1$. The present paper forms the
first step towards filling this lacuna in the theory.

The algebra-geometric approach we employ in the present paper to solve
the Q1 lattice equation \eqref{eq:1.1} uses a variant of the theory of finite-gap integration (see \cite{Matveev} and references therein)
based on the constuction of integrable symplectic maps in combination with the discrete version of the
Liouville-Arnold theory \cite{Quispel,Bruschi,Suris,Veselov,Veselov1}. In contrast to the approach presented in \cite{Cao,Cao1,Cao2}, which
employs associated completely integrable Hamiltonian systems (in terms of an auxiliary continuous time-variable)
for the integration of the lattice systems, in the present approach we circumvent the need for an associated
continuous variable and use the discrete equation itself as the starting point. This, through the property
of multidimensional consistency, cf. \cite{Hietarinta}, provides the Lax pair, the commuting matrix operators, as
well as the spectral curve associated with a hyperelliptic Riemann surface, in the details of the construction of
solutions.

This paper is organised as follows. In Section 2,  we present the relevant Lax pair as well as the underlying continuous
systems associated with the quadrliateral lattice through continuum limits.
In Section 3, based on two commuting operators, we obtain the relevant symplectic map, whose integrability is proved by
using the $r$-matrix and (quasi-) Abel-Jacobi variables. As a consequence, the discrete flows are constructed via the
iteration of integrable symplectic maps in Section 4. Moreover, the evolution of the potential along the discrete flows is
inverted with the help of the Baker-Akhiezer functions. As a result, the algebro-geometric integration of the Q1 lattice
equation is deduced by the commutativity of two discrete phase flows sharing the same invariants. We finish the paper with
some conclusive remarks in Section 5.

\section{Lax pairs and associated continuous equations}\setcounter{equation}{0}

By the method of multidimensional consistency, cf. \cite{Hietarinta}, a basic discrete spectral problem is derived directly from equation (\ref{eq:1.1})
\begin{equation}\label{eq:2.1}
\tilde{\chi}=\mathcal{D}^{(\beta)}(\lambda;b)\chi,\ \ \mathcal{D}^{(\beta)}(\lambda;b)=\displaystyle\frac{1}{B}\begin{pmatrix}
\lambda b& \lambda^{2}\delta^{2}\beta +\beta B^{2}\\
\beta &\lambda b
\end{pmatrix},
\end{equation}
where $B=(b^{2}-\delta^{2}\beta^{2})^{1/2}, \beta b=\tilde{u}-u$. The factor $B$ is there to avoid that non-trivial relations
arise from the determinantal condition of the zero-curvature relation, and one way of guaranteeing that happens is by requiring that
$\mathrm{det}\mathcal{D}^{(\beta)}(\lambda;b)=\lambda^{2}-\beta^{2}$ is a
constant. The general form \eqref{eq:2.1} of the Lax pair for quadrilateral lattice equations was given in \cite{Nijhoff0}, in the context of the one
for the generic case of the lattice KN system, i.e., Q4 lattice equation. The latter contains as a special case
the Q1 model, cf. also \cite{Adler2}. In the special case of the Q1 lattice equation (\ref{eq:1.1}) the zero-curvature representation adopts the form
\begin{align}\begin{split}\label{eq:2.2}
&\tilde{\chi}=\mathcal{D}^{(\beta_{1})}\chi \equiv \mathcal{D}^{(\beta_{1})}(\lambda;b_{1})\chi,\ \quad {\rm with}\quad b_{1}=(\tilde{u}-u)/\beta_{1},\\
&\bar{\chi}=\mathcal{D}^{(\beta_{2})}\chi \equiv \mathcal{D}^{(\beta_{2})}(\lambda;b_{2})\chi,\ \quad {\rm with} \quad b_{2}=(\bar{u}-u)/\beta_{2}.
\end{split}\end{align}
In fact, computing the difference $\bar{\tilde{\chi}}-\tilde{\bar{\chi}}$ yields the following identity:
\begin{equation}\label{eq:2.3}
\bar{\mathcal{D}}^{(\beta_{1})}\mathcal{D}^{(\beta_{2})}-\tilde{\mathcal{D}}^{(\beta_{2})}\mathcal{D}^{(\beta_{1})}=\displaystyle \frac{\Xi^{(0,2)}}{\bar{B}_{1}\tilde{B}_{2}B_{1}B_{2}}\begin{pmatrix}
\lambda^{2}S_{1}+\beta_{1}\beta_{2}\bar{B}_{1}\tilde{B}_{2}S_{0} & \lambda^{2}\delta^{2}S_{2}+\lambda S_{3}\\
\lambda S_{2} &\lambda^{2}S_{1}-\beta_{1}\beta_{2}B_{1}B_{2}S_{0}
\end{pmatrix},
\end{equation}
in which $B_{k}=(b_{k}^{2}-\delta^{2}\beta_{k}^{2})^{1/2}$, and where the quantities $S_j$ are given by
\begin{align*}
\begin{split}
&S_{0}=-\displaystyle\frac{\tilde{b}_{2}b_{2}+\bar{b}_{1}b_{1}-\delta^{2}(\beta_{1}^{2}+\beta_{2}^{2})}
{(\tilde{B}_{2}B_{2}+\bar{B}_{1}B_{1})\beta_{1}^{2}\beta_{2}^{2}},\\
&S_{1}=-\displaystyle\frac{\delta^{2}(\bar{\tilde{u}}-\tilde{u}-\bar{u}+u)(I_{1}J_{2}+I_{2}J_{1})}
{(\tilde{B}_{2}B_{1}I_{2}+\bar{B}_{1}B_{2}I_{1})\beta_{1}^{2}\beta_{2}^{2}},\\
&S_{2}=-\displaystyle\frac{(\bar{\tilde{u}}-\tilde{u}-\bar{u}+u)(I_{1}J_{2}+I_{2}J_{1})}
{(\tilde{B}_{2}B_{1}J_{2}+\bar{B}_{1}B_{2}J_{1})\beta_{1}^{2}\beta_{2}^{2}},\\
&S_{3}=\displaystyle\frac{\delta^{2}(\bar{\tilde{u}}-\tilde{u}-\bar{u}+u)}
{(\tilde{B}_{2}B_{1}K_{2}+\bar{B}_{1}B_{2}K_{1})\beta_{1}^{2}\beta_{2}^{2}}
[(I_{1}K_{2}+I_{2}K_{1})(\beta_{1}^{2}+\beta_{2}^{2})-(J_{1}K_{2}+J_{2}K_{1})(\bar{\tilde{u}}-u)],
\end{split}
\end{align*}
with
\begin{align*}
\begin{split}
&I_{1}=\tilde{b}_{2}b_{1}+\delta^{2}\beta_{1}\beta_{2},\ \ J_{1}=\beta_{1}\tilde{b}_{2}+\beta_{2}b_{1},\ \
K_{1}=\beta_{1}\tilde{b}_{2}B_{1}^{2}+\beta_{2}b_{1}\tilde{B}_{2}^{2},\\
&I_{2}=\bar{b}_{1}b_{2}+\delta^{2}\beta_{1}\beta_{2},\ \ J_{2}=\beta_{2}\bar{b}_{1}+\beta_{1}b_{2},\ \
K_{2}=\beta_{2}\bar{b}_{1}B_{2}^{2}+\beta_{1}b_{2}\bar{B}_{1}^{2}.
\end{split}
\end{align*}
From equation \eqref{eq:2.3} we conclude that the zero-curvature condition (implying $\bar{\tilde{\chi}}=\tilde{\bar{\chi}}$) is satisfied iff $\Xi^{(2,0)}=0$,
i.e. if the Q1 lattice equation \eqref{eq:1.1} holds for the function $u$, since generically at least one of the entries in the
matrix on the right hand side of \eqref{eq:2.3} is nonzero. Thus, the relation \eqref{eq:2.3} forms a bridge between the discrete zero-curvature equation and the Q1 model.

Let us now consider continuum limits of the Q1 equation. A special sequel of continuum limits yield the following form for a semi-continuum and
full continuum limit respectively,
\begin{align}\label{eq:2.4}
&\Xi^{(1,1)}\equiv\beta_{1}^{2}\tilde{u}_{x}u_{x}-(\tilde{u}-u)^{2}+\delta^{2}\beta_{1}^{4}=0,\\ \label{eq:2.5}
&\Xi^{(2,0)}\equiv u_{y}+\displaystyle \frac{1}{4}\big(u_{xxx}-\displaystyle \frac{3}{2}\displaystyle \frac{u_{xx}^{2}-4\delta^{2}}{u_{x}}\big)=0.
\end{align}
In fact, setting $\beta_{2}=c_{2}\varepsilon$, and $\bar{u}(x,y)=u(x+\beta_{2},y-\beta_{2}^{3}/3)$, then as remains $\beta_{1}$ constant while
$\beta_{2}\rightarrow 0$, i.e. $\varepsilon\rightarrow 0$, we have the Taylor expansion
\begin{equation*}
\Xi^{(0,2)}\equiv c_{2}^{2}\varepsilon^{2}\Xi^{(1,1)}+O(\varepsilon^{3}).
\end{equation*}
Furthermore, setting $\beta_{k}=c_{k}\varepsilon$, and $\tilde{u}(x,y)=u(x+\beta_{1},y-\beta_{1}^{3}/3)$, and
$\bar{u}(x,y)=u(x+\beta_{2},y-\beta_{2}^{3}/3)$. As $\varepsilon\rightarrow 0$, we obtain
\begin{equation*}
\Xi^{(0,2)}\equiv \displaystyle \frac{2}{3}u_{x}(c_{1}^{2}-c_{2}^{2})\varepsilon^{2}\Xi^{(2,0)}+O(\varepsilon^{3}).
\end{equation*}
We note that equation (\ref{eq:2.5}) is a special case of the Krichever-Novikov equation,\cite{Krichever}, which in its most general form
is given by
\begin{equation}\label{eq:2.6}
u_{y}=u_{xxx}-\displaystyle\frac{3}{2u_{x}}(u_{xx}^{2}-r(u))+cu_{x},\ \ r^{(5)}=0,
\end{equation}
with $r(u)$ an arbitrary quadratic polynomial with constant coefficients. V Adler, \cite{Adler1}, discovered the following B\"acklund transformation for
\eqref{eq:2.6}:
\begin{equation}\label{eq:2.7}
u_{x}v_{x}=h(u,v),
\end{equation}
where $h(u,v)$ is an arbitrary symmetric biquadratic polynomial, i.e., a polynomial of degree less than 3 in each variable. Furthermore, the nonlinear
superposition principle for (\ref{eq:2.7}), i.e., the prumtability condition of the B\"acklund transform, in the most general, elliptic, form gives
rise to the Q4 lattice equation, \cite{Adler1,Adler2}. In the more special case of the Q1 equation (\ref{eq:2.7}) coincides with the semicontinuous
equation \eqref{eq:2.4}) written as
\begin{equation}\label{eq:2.8}
u_{x}\tilde{u}_{x}=\beta_{1}^{-2}(\tilde{u}-u)^{2}-\delta^{2}\beta_{1}^{2},
\end{equation}
by identifying  $v=\tilde{u}$. The observation that the B\"acklund transformed quantity $\tilde{u}$ can be identified with a lattice shift goes back to
\cite{Levi,Levi1981}. On the level of the Lax pair,  the semi-discrete zero curvature equation  $\Lambda=\mathcal{D}^{(\beta)}_{x}- \tilde{\mathcal{U}} \mathcal{D}^{(\beta)}+\mathcal{D}^{(\beta)}\mathcal{ U}=0$, with $\mathcal{U}$ given by
\begin{equation}\label{eq:2.9}
\partial_{x}\chi=\mathcal{U}(\lambda;u)\chi=\displaystyle \frac{1}{\lambda u_{x}}\begin{pmatrix}0 & \lambda^{2}\delta^{2}+u_{x}^{2}\\ 1& 0\end{pmatrix}\chi,
\end{equation}
gives rise to the semi-discrete equation \eqref{eq:2.4} as comatibility condition.
Indeed, by direct calculations we have
\begin{equation*}
\Lambda=\displaystyle \frac{-\Xi^{(1,1)}}{\beta_{1} B_{1}\tilde{u}_{x}u_{x}}\Big[\displaystyle\frac{b_{1,x}}{B_{1}^{2}}
\begin{pmatrix}\lambda \delta^{2}\beta_{1} & \lambda^{2}\delta^{2}b_{1}\\ b_{1}& \lambda \delta^{2}\beta_{1} \end{pmatrix}+
\displaystyle\frac{1}{\lambda}\begin{pmatrix}\tilde{u}_{x} &0\\0&-u_{x} \end{pmatrix}\Big].
\end{equation*}
We mention in passing that the  Krichever-Novikov equation (\ref{eq:2.6}) arose in \cite{Krichever} from the problem
of finding special finite-gap solutions of the Kadomtsev-Petviashvili equation associated with rank 2 holomorphic vector bundles.
We mention also that already early on it was observed that the semi-discrete system is perfectly adapted for performing numerical calculations,
and for producing the plots of solutions, which reveals that the algebro-geometric solutions have a practical
value as standard special functions as well \cite{Bobenko,Belokolos,Bobenko1989}.

Motivated by these known facts, this paper is dedicated to the problem of constructing the algebro-geometric integration for the Q1 equation
(\ref{eq:1.1}). To set up the finite-gap scheme for Q1 it is most convenient to perform the calculations from a Lax representation perspective. In the development of this method, a fundamental role is played by the Burchnall-Chaundy theory of commuting differential operators \cite{Burchnall,Burchnall1928}, whose discrete analogue concerning commutative rings of periodic difference
operators was developed \cite{Naiman,Glazman}. Their common eigenfunctions are vector bundles over a Riemann surface defined by the corresponding eigenvalues,
and this forms the underlying geometry for the reconstruction of the potentials from a special class of spectral
functions which themselves can be obtained from solving a classical Jacobi inversion problem for Abelian integrals on hyperelliptic Riemann surfaces \cite{Akhiezer,Its1975,Kri1977}.
Furthermore, the algebra-geometric solutions of the associated discrete systems can be expressed in terms of the Riemann theta functions associated with
the Riemann surface, and a suitable choice of a homology basis of curves (see \cite{Akh1960,Kac,Dubrovin1976,Date,Flaschka,Mumford1979,Krichever2003,Gesztesy,Cao,Geng} and the reference therein).
The approach in this paper for the solution of the discrete equation is based on the construction of commuting integrable symplectic (dynamical) maps,
which in turn can be resolved in terms of the algebro-geometric data.

\section{Construction of the nonlinear integrable symplectic map}\setcounter{equation}{0}

In this section we discuss how a linear map can be nonlinearized to produce a nonlinear integrable symplectic map
on the symplectic manifold $\mathcal{N}=(\mathbb{R}^{2N},\mathrm{d}p\wedge \mathrm{d}q)$ with associated
coordinates $(p,q)=(p_{1},\ldots,p_{N},q_{1},\ldots,q_{N})^{T}$, where $N$ is a positive integer. A map $
\mathcal{S}: (p,q)\mapsto (\tilde{p},\tilde{q})$ is a symplectic transformation \cite{Arnold,Goldstein,Jordan}, if
$\mathcal{S}^{*}(\mathrm{d}p\wedge \mathrm{d}q)=\mathrm{d}p\wedge \mathrm{d}q$. A well-defined notion of
integrability for symplectic maps was first given by Veselov \cite{Veselov,Veselov1}. It is entirely analogous to
that of the Liouville-Arnold in the continuous-time case, i.e., there exists $N$ smooth functions $I_{1},
\ldots,I_{N}:\mathbb{R}^{2N}\rightarrow\mathbb{R}$ with the following properties:

1. The functions are invariants of the map, i.e., $\mathcal{S}^{*}I_{j}=I_{j}$.

2. The invariant functions are in involution with respect to the Poisson bracket, i.e., $\{I_{i},I_{j}\}=0$.

3. The invariant functions are functionally independent throughout the phase space.

\noindent To construct an integrable nonlinear symplectic map suitable for the integration of the
Q1 equation, we use the nonlinearisation method of \cite{Cao,Cao1,Cao2} and apply it to the linear
map associated with the spectral problem (\ref{eq:2.1}) of Q1, which is given by:
\begin{equation}\label{eq:3.1}
\quad \begin{pmatrix}
\tilde{p}_{j}\\
\tilde{q}_{j}
\end{pmatrix}=
(\alpha_{j}^{2}-\beta^{2})^{-1/2}\mathcal{D}^{(\beta)}(\alpha_{j};b)
\begin{pmatrix}
 p_{j}\\
 q_{j}
\end{pmatrix}, \ \ 1\leq j\leq N,
\end{equation}
where $\{\alpha_{j}\}_{j=1}^{N}$ are the corresponding eigenvalues. We assume, moreover, that these
eigenvalues $\alpha_{1}^{2},\ldots,\alpha_{N}^{2}$ are mutually distinct and non-zero.
It turns out that the linear map (\ref{eq:3.1}) can be extended to a nonlinear integrable symplectic
map by imposing a restriction on the discrete potential\footnote{This assertion is based on the
observation, going back to \cite{Krichever}, that there exists an operator commuting with a member in the Lax representation for nonlinear
equations of Korteweg-de Vries type, when discussing the associated finite gap classes of exact solutions.}
which is the quantity $b$.
This fact has led in several cases to the construction of integrable symplectic maps in the process of
constructing the finite gap solutions for several discrete soliton equations \cite{Cao,Cao1,Cao2}.
In the present case, the operator associated with the
Lax pair (\ref{eq:2.2}) is the Darboux matrix $\mathcal{D}^{(\beta)}(\lambda;b)$ given by (\ref{eq:2.1}). Through direct computation, we find that there is a matrix operator, i.e., the Lax matrix
\begin{equation}\label{eq:3.2}
\mathcal{L}(\lambda;p,q)=\begin{pmatrix}0& \delta^{2}(\lambda^{2}-<Aq,q>)\\1&0\end{pmatrix}+\begin{pmatrix} \lambda Q_{\lambda}(p,q)& -Q_{\lambda}(Ap,p)
\\ Q_{\lambda}(Aq,q)& -\lambda Q_{\lambda}(p,q)\end{pmatrix},
\end{equation}
that commutes with $\mathcal{D}^{(\beta)}(\lambda;b)$:
\begin{equation}\label{eq:3.3}
\mathcal{L}(\lambda;\tilde{p},\tilde{q})\mathcal{D}^{(\beta)}(\lambda;b)-\mathcal{D}^{(\beta)}(\lambda;b)\mathcal{L}(\lambda;p,q)=0,
\end{equation}
where $A=\mathrm{diag}(\alpha_{1},\ldots,\alpha_{N}), <\xi,\eta>=\Sigma_{j=1}^{N}\xi_{j}\eta_{j}$ and $Q_{\lambda}(\xi,\eta)=<(\lambda^{2}-A^{2})^{-1}\xi,\eta>$. This commutative relation (\ref{eq:3.3}) in cases considered implies a quadratic equation providing the constraints on discrete potentials \cite{Cao,Cao1,Cao2}.

\noindent \textbf{Proposition 3.1.} The constraint for the discrete potential $b$ in the spectral problem (\ref{eq:2.1}) satisfies
\begin{equation}\label{eq:3.4}
P^{(\beta)}(b; p,q)\equiv b^{2}\mathcal{L}^{21}(\beta;p,q)+2b\mathcal{L}^{11}(\beta;p,q)-\mathcal{L}^{12}(\beta;p,q)=0.
\end{equation}
\noindent \emph{Proof.} First, note that
the Lax matrix (\ref{eq:3.2}) can be rewritten as
\begin{align*}
\mathcal{L}(\lambda;p,q)=\begin{pmatrix}0& \delta^{2}(\lambda^{2}-<Aq,q>)\\1&0\end{pmatrix}+\displaystyle\frac{1}{2}\sum_{j=1}^{N}
\big(\displaystyle\frac{\varepsilon_{j}}{\lambda-\alpha_{j}}+\displaystyle\frac{\sigma_{3}\varepsilon_{j}\sigma_{3}}{\lambda+\alpha_{j}}\big),
\ \ \varepsilon_{j}=\begin{pmatrix}p_{j}q_{j} &
-p_{j}^{2}\\q_{j}^{2} & -p_{j}q_{j}\end{pmatrix}.
\end{align*}
Then we calculate $\mathcal{L}(\lambda;\tilde{p},\tilde{q})\mathcal{D}^{(\beta)}(\lambda;b)-\mathcal{D}^{(\beta)}(\lambda;b)\mathcal{L}(\lambda;p,q)=I+II$,
\begin{align*}
I&=\begin{pmatrix}\displaystyle\frac{-\beta}{B}(\delta^{2}<A\tilde{q},\tilde{q}>+B^{2})& -\lambda\delta^{2}(<A\tilde{q},\tilde{q}>-<Aq,q>)\\0&\displaystyle\frac{\beta}{B}(\delta^{2}<Aq,q>+B^{2})\end{pmatrix},\\
II&=\displaystyle\frac{1}{2}\sum_{j=1}^{N}\displaystyle\frac{\tilde{\varepsilon}_{j}
\mathcal{D}^{(\beta)}(\lambda;b)-\mathcal{D}^{(\beta)}(\lambda;b)\varepsilon_{j}}{\lambda-\alpha_{j}}+
\displaystyle\frac{\sigma_{3}\tilde{\varepsilon}_{j}\sigma_{3}
\mathcal{D}^{(\beta)}(\lambda;b)-\mathcal{D}^{(\beta)}(\lambda;b)\sigma_{3}\varepsilon_{j}\sigma_{3}}{\lambda+\alpha_{j}}\\
&=\displaystyle\frac{b}{B}(<\tilde{p},\tilde{q}>-<p,q>)\sigma_{3}+\displaystyle\frac{\beta\delta^{2}}{B}
\begin{pmatrix}-<Aq,q>&\lambda(<\tilde{p},\tilde{q}>+<p,q>)\\
0&<A\tilde{q},\tilde{q}>\end{pmatrix},
\end{align*}
where we use $\tilde{\varepsilon}_{j}\mathcal{D}^{(\beta)}(\alpha_{j};b)=\mathcal{D}^{(\beta)}(\alpha_{j};b)\varepsilon_{j},\ \ \sigma_{3}\mathcal{D}^{(\beta)}(\alpha_{j};b)\sigma_{3}=-\mathcal{D}^{(\beta)}(-\alpha_{j};b)$. From equations (\ref{eq:2.1}) and (\ref{eq:3.1}), we have
\begin{align*}
&\tilde{p}=(A^{2}-\beta^{2})^{-1/2}\displaystyle\frac{1}{B}(bAp+\beta\delta^{2}A^{2}q+\beta B^{2}q),\\
&\tilde{q}=(A^{2}-\beta^{2})^{-1/2}\displaystyle\frac{1}{B}(\beta p+bAq).
\end{align*}
Hence
\begin{align*}
\mathcal{L}(\lambda;\tilde{p},\tilde{q})\mathcal{D}^{(\beta)}(\lambda;b)-\mathcal{D}^{(\beta)}(\lambda;b)\mathcal{L}(\lambda;p,q)&=
\displaystyle\frac{1}{B}[b(<\tilde{p},\tilde{q}>-<p,q>)-\\
&\beta\delta^{2}(<A\tilde{q},\tilde{q}>+<Aq,q>)-\beta B^{2}]\sigma_{3}\\
&=-\displaystyle\frac{\beta}{B}P^{(\beta)}(b;p,q)\sigma_{3},
\end{align*}
where $\sigma_{3}$ is the usual Pauli matrix. This concludes the proof by using equation \eqref{eq:3.3}.  \hfill $\Box$

\noindent Following the spirit of \cite{Krichever}, we use the Burchnall-Chaundy theory \cite{Burchnall,Burchnall1928,Naiman,Glazman} to study the
commuting operators $\mathcal{L}(\lambda;p,q)$ and $\mathcal{D}^{(\beta)}(\lambda;b)$. We shall investigate their
common eigenfunctions and the corresponding eigenvalues which lie on a Riemann surface. In our case, the operator $
\mathcal{L}(\lambda;p,q)$ has two eigenvalues,
\begin{equation}\label{eq:3.5}
\pm \mathcal{H}_{\lambda}=\pm\sqrt{-\mathcal{F}_{\lambda}},
\end{equation}
where
\begin{align}\label{eq:3.6}
\begin{split}
\mathcal{F}_{\lambda}=&\mathcal{F}_{\lambda}(p,q)\overset{\triangle}{=}\mathrm{det}\mathcal{L}(\lambda;p,q)\\
=&(-\delta^{2}\lambda^{2}+\delta^{2}<Aq,q>+Q_{\lambda}(Ap,p))(1+Q_{\lambda}(Aq,q))-\lambda^{2}Q_{\lambda}^{2}(p,q)\\
=&-\delta^{2}\lambda^{2}+Q_{\lambda}(Ap,p)-\delta^{2}Q_{\lambda}(A^{3}q,q)+\delta^{2}<Aq,q>Q_{\lambda}(Aq,q)+\\
&+Q_{\lambda}(Ap,p)Q_{\lambda}(Aq,q)-\lambda^{2}Q_{\lambda}^{2}(p,q),
\end{split}
\end{align}
with $Q_{\lambda}(A^{3}q,q)=\lambda^{2}Q_{\lambda}(Aq,q)-<Aq,q>$. Moreover, $\mathcal{F}_{\lambda}$ is a rational function of $\zeta=\lambda^{2}$, and has simple pole at each point $\alpha_{j}^{2}$. Inspired by the relation between the eigenvalues $\mathcal{H}_{\lambda}$ and $\lambda$, gienv by (\ref{eq:3.5}), we consider the factorization of $\mathcal{F}_{\lambda}$
\begin{equation}\label{eq:3.7}
\mathcal{F}_{\lambda}=-\delta^{2}\displaystyle\frac{\prod_{j=1}^{N+1}(\zeta-\lambda_{j}^{2})}{\alpha(\zeta)}
=-\delta^{2}\displaystyle\frac{R(\zeta)}{\zeta\alpha^{2}(\zeta)},\ \ \alpha(\zeta)=\Pi_{k=1}^{N}(\zeta-\alpha_{k}^{2}),
\end{equation}
and then construct a hyperelliptic spectral curve associated with a 2-sheeted Riemann surface of genus $g=N$,  cf. \cite{Farkas,Griffiths,Mumford},
\begin{equation}\label{eq:3.8}
\mathcal{R}: \xi^{2}=R(\zeta).
\end{equation}
Since $\mathrm{deg}R=2N+2$, the above curve (\ref{eq:3.8}) has two infinities $\infty_{+}$, $\infty_{-}$. For any $\zeta\in \mathbb{C}$, in the non-branch case (not equal to $\lambda_{j}^{2}, \alpha_{j}^{2}$ or $ 0$) we call the collection of points on $\mathcal {R}$ of the form
\begin{equation*}
\mathfrak{p}(\zeta)=\big(\zeta,\xi=\sqrt{R(\zeta)}\big),\ \
(\tau\mathfrak{p})\big(\zeta)=(\zeta,\xi=-\sqrt{R(\zeta)}\big),
\end{equation*}
the upper and lower sheets, respectively, where $\tau:\mathcal{R}\rightarrow\mathcal{R}$ is the map of changing sheets. The branch point, given by $\zeta = 0$ and $\xi = 0$, is denoted by $\mathfrak{0}$.

\noindent  To proceed with the actual integration of the map, we first list some basic objects on $\mathcal{R}$
\cite{Farkas,Griffiths,Mumford}. Let $a_{1},\cdots,a_{g},b_{1},\cdots,b_{g}$ be the canonical basis of the homology
group $H_{1}(\mathcal{R})$ and $\omega_{1}^{\prime},\cdots,\omega_{g}^{\prime}$ be the basis of holomorphic differentials:
\begin{equation}\label{eq:3.9}
\omega_{l}^{\prime}=\sum_{s=1}^{g}\displaystyle\frac{\zeta^{g-s}\mathrm{d}\zeta}{2\sqrt{R(\zeta)}},\ \ 1\leq l\leq g,
\end{equation}
whose integral along $a_{k}$ is denoted by $a_{lk}$. Then (\ref{eq:3.9}) in the vector form $\vec{\omega}^\prime=(\omega_1^\prime,\cdots,\omega_g^\prime)^T$ can be normalized into $\vec{\omega}=(\omega_{1},\cdots,\omega_{g})^{T}=C\vec{\omega}^\prime$ with $C=(a_{lk})^{-1}_{g\times g}$.

\noindent Periodic vectors  $\vec\delta_k,\,\vec B_k$ are defined as integrals of $\vec\omega$ along $a_k,\,b_k$, respectively. They span a lattice $\mathscr T$, which defines the Jacobian variety $J(\mathcal R)=\mathbb C^g/\mathscr T$. The matrix $B = (\vec B_1, \ldots , \vec B_g)$ is used to construct the Riemann theta functions which will be used in Section 4,
\begin{equation}\label{eq:3.10}
\theta(z,B)=\sum_{z^{\prime}\in \mathbb{Z}^{g}}\exp\pi\sqrt{-1}(<Bz^{\prime},z^{\prime}>+2<z,z^{\prime}>),\ \ z\in \mathbb{C}^{g}.
\end{equation}
The Abel map $\mathscr A: \mathrm{Div}(\mathcal{R})\rightarrow J(\mathcal{R})$ defined as
\begin{equation}\label{eq:3.11}
\mathscr A(\mathfrak p)=\int_{\mathfrak{p}_0}^{\mathfrak{p}}\vec\omega,\quad \mathscr A(\Sigma n_{k}\mathfrak{p}_k)=\Sigma n_{k}\mathscr A(\mathfrak{p}_k),
\end{equation}
is the key ingredient in the Jacobi inversion problem.

\noindent As a consequence we obtain a nonlinear map arising from the linear map (\ref{eq:3.1}),
\begin{equation}\label{eq:3.12}
\mathcal{S}_{\beta}: \ \ \begin{pmatrix}\tilde{p} \\ \tilde{q}\end{pmatrix}=(A^{2}-\beta^{2})^{-1/2}\displaystyle\frac{1}{B}
\begin{pmatrix}bAp+\beta\delta^{2}A^{2}q+\beta B^{2}q\\ \beta p+bAq\end{pmatrix}\Bigg|_{b=f_{\beta}(p,q)},
\end{equation}
where $b=f_{\beta}(p,q)$ is given by the roots of quadratic equation (\ref{eq:3.4}),
\begin{equation}\label{eq:3.13}
b=f_{\beta}(p,q)=\displaystyle
\frac{1}{1+Q_{\beta}(Aq,q)}\big(-\beta Q_{\beta}(p,q)\pm \mathcal{H}_{\beta}),
\end{equation}
which is single-valued as a function of $\mathfrak{p}(\beta^{2})\in \mathcal{R}$. Actually, $\beta b$ are the values of the following meromorphic function on the curve $\mathcal{R}$:
\begin{equation*}
\mathfrak{b}(\mathfrak{p})=\displaystyle
\frac{1}{1+Q_{\beta}(Aq,q)}\big(-\beta^{2} Q_{\beta}(p,q)+\frac{\delta\xi}{\alpha(\beta^{2})}\big),
\end{equation*}
at the points $\mathfrak{p}(\beta^{2})$ and $(\tau\mathfrak{p})(\beta^{2})$, respectively. Hence, the nonlinear map $\mathcal{S}_{\beta}$ is well-defined.

The next step is to show that $\mathcal{S}_{\beta}$ given by (\ref{eq:3.12}) is an integrable symplectic map on $\mathcal{N}=(\mathbb{R}^{2N},\mathrm{d}p\wedge \mathrm{d}q)$.
In order to get the symplecticity of $\mathcal{S}_{\beta}$, we calculate
\begin{equation}\label{eq:3.14}
\begin{pmatrix}
\mathrm{d}\tilde{p}_{j}\\
\mathrm{d}\tilde{q}_{j}
\end{pmatrix}=
(\alpha_{j}^{2}-\beta^{2})^{-1/2}\Big(\mathcal{D}^{(\beta)}(\alpha_{j};b)
\begin{pmatrix}
\mathrm{d} p_{j}\\
 \mathrm{d}q_{j}
\end{pmatrix}+\mathcal{C}^{(\beta)}(\alpha_{j};b)\begin{pmatrix}
\mathrm{d} p_{j}\\
 \mathrm{d}q_{j}
\end{pmatrix}\mathrm{d}b\Big), \ \ 1\leq j\leq N,
\end{equation}
where
\begin{align*}
\mathcal{C}^{(\beta)}(\alpha_{j};b)=\displaystyle\frac{\mathrm{d}}{\mathrm{d}b}\mathcal{D}^{(\beta)}(\alpha_{j};b)=
-\displaystyle\frac{b}{B^{2}}\mathcal{D}^{(\beta)}(\alpha_{j};b)+\displaystyle\frac{1}{B}\begin{pmatrix}
\alpha_{j}&2\beta b\\
0&\alpha_{j}
\end{pmatrix},
\end{align*}
by using equation (\ref{eq:3.12}). Then,
\begin{equation}\label{eq:3.15}
\sum\limits_{j=1}^{N} (\mathrm{d}\tilde{p}_{j} \wedge
\mathrm{d}\tilde{q}_{j}-\mathrm{d}p_{j} \wedge\mathrm{d}
q_{j})=\displaystyle\frac{\beta}{2B^{2}}\mathrm{d}P^{(\beta)}(b;p,q) \wedge
\mathrm{d}b,
\end{equation}
which implies $\mathcal{S}_{\beta}^{*}(\mathrm{d}p\wedge \mathrm{d}q)=\mathrm{d}\tilde{p}\wedge \mathrm{d}\tilde{q}
=\mathrm{d}p\wedge \mathrm{d}q$, since $b=f_{\beta}(p,q)$ given by equation (\ref{eq:3.13}) satisfies the quadratic
equation (\ref{eq:3.4}). Hence, $\mathcal{S}_{\beta}$ is a symplectic map on $\mathcal{N}=(\mathbb{R}^{2N},
\mathrm{d}p\wedge \mathrm{d}q)$.

Having asserted the symplecticity of the map, we now turn to the construction of its invariant. In fact,
by taking the determinant on equation (\ref{eq:3.3}), we have $\mathrm{det}\mathcal{L}(\lambda;\tilde{p},
\tilde{q})=\mathrm{det}\mathcal{L}(\lambda;p,q)$, i.e.,
\begin{equation}\label{eq:3.16}
\mathcal{\tilde{F}}_{\lambda}=\mathcal{F}_{\lambda}.
\end{equation}
Thus, invariants of the map $\mathcal{S}_{\beta}$ can be generated by the partial fraction expansion,
\begin{equation}\label{eq:3.17}
\mathcal{F}_{\lambda}=-\delta^{2}\lambda^{2}+\sum_{k=1}^{N}\displaystyle\frac{E_{k}}{\lambda^{2}-\alpha_{k}^{2}},
\end{equation}
where
\begin{align}\label{eq:3.18}
\begin{split}
E_{k}=&\alpha_{k}p_{k}^{2}-\delta^{2}\alpha_{k}^{3}q_{k}^{2}+\delta^{2}<Aq,q>\alpha_{k}q_{k}^{2}-p_{k}^{2}q_{k}^{2}\\
&+\displaystyle\frac{\alpha_{k}}{2}\sum\limits_{1\leq j\leq N;j\neq k}\displaystyle\frac{(p_{k}q_{j}-p_{j}q_{k})^{2}}{\alpha_{k}^{2}-\alpha_{j}^{2}}-
\displaystyle\frac{(p_{k}q_{j}+p_{j}q_{k})^{2}}{\alpha_{k}^{2}+\alpha_{j}^{2}}.
\end{split}
\end{align}
In fact, substituting (\ref{eq:3.17}) into the both sides of equation (\ref{eq:3.16}) and comparing the residues at $\alpha_{k}^{2}$, we obtain
\begin{equation}\label{eq:3.19}
\tilde{E}_{k}=E_{k},\ \ 1\leq k\leq N.
\end{equation}

 The next step is to show that $N$ invariant functions $E_{1},\ldots,E_{N}$ on the phase space
$\mathbb{R}^{2N}$ for the symplectic map $\mathcal{S}_{\beta}$ are in involution with respect to the symplectic
structure, and are functionally independent. These conditions are essential conditions to assert the
Liouville integrability of symplectic maps. We employ a different approach here from
previous papers, \cite{Cao,Cao1,Cao2}, by employing an $r$-matrix structure to exhibit the validity of
these conditions. The relevant $r$-matrix structure is similar to the well known
cases \cite{Faddeev,Babelon,Gerdjikov}, and through direct calculation we find that the Lax matrix obeys the
fundamental Poisson bracket
\begin{equation}\label{eq:3.20}
\{\mathcal{L}(\lambda)\underset{,}\otimes \mathcal{L}(\mu)\} =
[r(\lambda,\mu),\mathcal{L}_{1}(\lambda)]+[r^{\prime}(\lambda,\mu),\mathcal{L}_{2}(\mu)],
\end{equation}
where $\mathcal{L}(\lambda;p,q)$ is often written as $\mathcal{L}(\lambda)$ for short, $\mathcal{L}_{1}(\lambda)=\mathcal{L}(\lambda)\otimes I$, $\mathcal{L}_{2}(\mu)=I\otimes \mathcal{L}(\mu)$, and $r^{\prime}(\lambda,\mu)=-r(\mu,\lambda)$ satisfies
\begin{align}\label{eq:3.21}
\begin{split}
r(\lambda,\mu)&=\displaystyle\frac{1}{\lambda^{2}-\mu^{2}}\big(\lambda(\sigma_{1}\otimes\sigma_{1}+\sigma_{2}\otimes\sigma_{2})+
\mu(\sigma_{3}\otimes\sigma_{3}+I)\big)+2\delta^{2}\lambda\sigma_{+}\otimes\sigma_{+}\\
&=\displaystyle\frac{2}{\lambda^{2}-\mu^{2}}\begin{pmatrix}\mu
& 0&0&0 \\0&0&\lambda&0\\0&\lambda&0&0\\0&0&0&\mu
\end{pmatrix}+2\delta^{2}\lambda\begin{pmatrix}0
& 0&0&1 \\0&0&0&0\\0&0&0&0\\0&0&0&0
\end{pmatrix},
\end{split}
\end{align}
with $\sigma_{1},\sigma_{2},\sigma_{3}, \sigma_{+}$ the Pauli matrices and $I$ the usual unit matrix.

\noindent In addition to \eqref{eq:3.20}, there are two further matrix functions $s,s^{\prime}$ such that
\begin{equation}\label{eq:3.22}
\{\mathcal{L}^{2}(\lambda)\underset{,}\otimes \mathcal{L}^{2}(\mu)\} =
[s,\mathcal{L}_{1}(\lambda)]+[s^{\prime},\mathcal{L}_{2}(\mu)].
\end{equation}
This is derived from
\begin{align}\label{eq:3.23}
\begin{split}
\{\mathcal{L}^{2}(\lambda)\underset{,}\otimes \mathcal{L}^{2}(\mu)\}=&\mathcal{L}_{1}(\lambda)\mathcal{L}_{2}(\mu)\{\mathcal{L}(\lambda)\underset{,}\otimes \mathcal{L}(\mu)\}+\\
&\mathcal{L}_{1}(\lambda)\{\mathcal{L}(\lambda)\underset{,}\otimes \mathcal{L}(\mu)\}\mathcal{L}_{2}(\mu)+\\
&\mathcal{L}_{2}(\mu)\{\mathcal{L}(\lambda)\underset{,}\otimes \mathcal{L}(\mu)\}\mathcal{L}_{1}(\lambda)+\\
&\{\mathcal{L}(\lambda)\underset{,}\otimes \mathcal{L}(\mu)\}\mathcal{L}_{2}(\mu)\mathcal{L}_{1}(\lambda).
\end{split}
\end{align}
Substituting (\ref{eq:3.20}) into (\ref{eq:3.23}) and by using the formula $\mathcal{L}_{1}(\lambda)\mathcal{L}_{2}(\mu)=\mathcal{L}_{2}(\mu)\mathcal{L}_{1}(\lambda)=\mathcal{L}(\lambda)\otimes \mathcal{L}(\mu)$, we get
\begin{align*}
\begin{split}
&s=\mathcal{L}_{1}(\lambda)\mathcal{L}_{2}(\mu)r(\lambda,\mu)+\mathcal{L}_{1}(\lambda)r(\lambda,\mu)\mathcal{L}_{2}(\mu)
+\mathcal{L}_{2}(\mu)r(\lambda,\mu)\mathcal{L}_{1}(\lambda)+r(\lambda,\mu)\mathcal{L}_{2}(\mu)\mathcal{L}_{1}(\lambda),\\
&s^{\prime}=\mathcal{L}_{1}(\lambda)\mathcal{L}_{2}(\mu)r^{\prime}(\lambda,\mu)+\mathcal{L}_{1}(\lambda)r^{\prime}(\lambda,\mu)\mathcal{L}_{2}(\mu)+
\mathcal{L}_{2}(\mu)r^{\prime}(\lambda,\mu)\mathcal{L}_{1}(\lambda)+r^{\prime}(\lambda,\mu)\mathcal{L}_{2}(\mu)\mathcal{L}_{1}(\lambda).
\end{split}
\end{align*}
As a consequence of the $r$-matrix structure we have the following:

\noindent
\textbf{Proposition 3.2.} The invariants $E_{1},\ldots,E_{N}$ of the symplectic map $\mathcal{S}_{\beta}$ are in pairwise involution.

\noindent \emph{Proof.} Since $\mathcal{L}^{2}(\lambda)=-\mathcal{F}_{\lambda}I, \mathcal{L}^{2}(\mu)=-\mathcal{F}_{\mu}I$, we calculate
\begin{equation}\label{eq:3.24}
\{\mathcal{F}_{\lambda},\mathcal{F}_{\mu}\}=\displaystyle\frac{1}{4}\mathrm{tr}\{\mathcal{L}^{2}(\lambda)\underset{,}\otimes \mathcal{L}^{2}(\mu)\}.
\end{equation}
Hence by equation (\ref{eq:3.22}), we obtain
\begin{equation}\label{eq:3.25}
\{\mathcal{F}_{\lambda},\mathcal{F}_{\mu}\}= 0, \ \ \forall \lambda,\mu \in\mathbb{C}.
\end{equation}
Substitute the partial fraction expansion (\ref{eq:3.17}) into (\ref{eq:3.25}), then calculate the residues, we have
\begin{equation}\label{eq:3.26}
\{E_{k}, E_{j}\} = 0, \ \ 1\leq j,k\leq N,
\end{equation}
which implies $\{E_{k}\}$ given by equation (\ref{eq:3.18}) are in involution.  \hfill $\Box$

Interestingly, by the $r$-matrix method, we can get the evolution of the Lax matrix along a phase flow resulting in the independence
for invariant functions \cite{Cao,Cao3}. In order to do this we calculate
\begin{equation}\label{eq:3.27}
\displaystyle\frac{\mathrm{d}}{\mathrm{
d}t_{\lambda}}\begin{pmatrix}p_{j}\\q_{j}\end
{pmatrix}=\begin{pmatrix}-\partial \mathcal{F}_{\lambda}/\partial q_{j}\\
\partial \mathcal{F}_{\lambda}/\partial p_{j}\end
{pmatrix}=\mathcal{W}(\lambda,\alpha_{j})\begin{pmatrix}p_{j}\\q_{j}\end
{pmatrix},
\end{equation}
where $t_{\lambda}$ is the flow variable corresponding to the Hamiltonian function $\mathcal{F}_{\lambda}$, then we obtain
\begin{equation}\label{eq:3.28}
\mathcal{W}(\lambda,\mu)=\displaystyle\frac{2}{
\lambda^{2}-\mu^{2}}\begin{pmatrix}\lambda \mathcal{L}^{11}(\lambda)&\mu \mathcal{L}^{12}(\lambda)\\
 \mu \mathcal{L}^{21}(\lambda)&-\lambda \mathcal{L}^{11}(\lambda)\end{pmatrix}
-2\delta^{2}\mu \mathcal{L}^{21}(\lambda)\sigma_{+}.
\end{equation}
 This statement can be cast in Lax form as follows:

\noindent
\textbf{Lemma 3.1.} The Lax matrix $\mathcal{L}(\mu)$ satisfies the evolution equation along the $t_{\lambda}$-flow,
\begin{equation}\label{eq:3.29}
\mathrm{d}\mathcal{L}(\mu)/\mathrm{d}t_{\lambda}=[\mathcal{W}(\lambda,\mu),\mathcal{L}(\mu)].
\end{equation}
\emph{Proof.} Since $\mathcal{L}^{2}(\lambda)=-\mathcal{F}_{\lambda}I$, we obtain
\begin{align}\label{eq:3.30}
\begin{split}
\{\mathcal{L}^{2}(\lambda)\underset{,}\otimes \mathcal{L}(\mu)\}&=\{-\mathcal{F}_{\lambda}I\underset{,}\otimes \mathcal{L}(\mu)\}\\
&=\begin{pmatrix}-\{\mathcal{F}_{\lambda},\mathcal{L}(\mu)\}&0\\0&-\{\mathcal{F}_{\lambda},\mathcal{L}(\mu)\}\end{pmatrix}\\
&=\begin{pmatrix}\mathrm{d}\mathcal{L}(\mu)/\mathrm{d}t_{\lambda}&0\\0&\mathrm{d}\mathcal{L}(\mu)/\mathrm{d}t_{\lambda}\end{pmatrix}.
\end{split}
\end{align}
By equation (\ref{eq:3.20}), we calculate the left hand side of (\ref{eq:3.30}) again and get
\begin{align}\label{eq:3.31}
\begin{split}
\{\mathcal{L}^{2}(\lambda)\underset{,}\otimes \mathcal{L}(\mu)\}=&\mathcal{L}_{1}(\lambda)\{\mathcal{L}(\lambda)\underset{,}\otimes \mathcal{L}(\mu)\} +\{\mathcal{L}(\lambda)\underset{,}\otimes \mathcal{L}(\mu)\}\mathcal{L}_{1}(\lambda)\\
=&\mathcal{L}_{1}(\lambda)r^{\prime}(\lambda,\mu)\mathcal{L}_{2}(\mu)-\mathcal{L}_{1}(\lambda)\mathcal{L}_{2}(\mu)r^{\prime}(\lambda,\mu)+\\
&r^{\prime}(\lambda,\mu)\mathcal{L}_{2}(\mu)\mathcal{L}_{1}(\lambda)-\mathcal{L}_{2}(\mu)r^{\prime}(\lambda,\mu)\mathcal{L}_{1}(\lambda)\\
=&[\mathcal{L}_{1}(\lambda)r^{\prime}(\lambda,\mu)+r^{\prime}(\lambda,\mu)\mathcal{L}_{1}(\lambda),\mathcal{L}_{2}(\mu)]\\
=&\begin{pmatrix}[\mathcal{W}(\lambda,\mu),\mathcal{L}(\mu)]&0\\0&[\mathcal{W}(\lambda,\mu),\mathcal{L}(\mu)]\end{pmatrix}.
\end{split}
\end{align}
Then comparing (\ref{eq:3.30}) and (\ref{eq:3.31}), equation (\ref{eq:3.29}) is verified. \hfill $\Box$

 We next address the problem of parametrising the solutions. In fact, by introducing an elliptic
(curvilinear) coordinate system $\{\nu_{j}^{2}\}$ defined by the zeros of the following function \cite{Arnold,Lame}:
\begin{equation}\label{eq:3.32}
\mathcal{L}^{21}(\lambda)=1+\sum_{j=1}^{g}\displaystyle\frac{\alpha_{j}q_{j}^{2}}{\lambda^{2}-\alpha_{j}^{2}}=\displaystyle\frac{\mathfrak{n}(\zeta)}{\alpha(\zeta)},\ \ \mathfrak{n}(\zeta)=\prod_{j=1}^{g}(\zeta-\nu_{j}^{2}),
\end{equation}
we consider one component of the equation (\ref{eq:3.29}),
\begin{equation}\label{eq:3.33}
\mathrm{d}\mathcal{L}^{21}(\mu)/\mathrm{d}t_{\lambda}=2(\mathcal{W}^{21}(\lambda,\mu)\mathcal{L}^{11}(\mu)-\mathcal{W}^{11}(\lambda,\mu)\mathcal{L}^{21}(\mu)),
\end{equation}
at points $\mu=\nu_{k}, 1\leq k\leq g$. Then the Dubrovin equations for our case 
\cite{Dubrovin,Gesztesy} is obtained\footnote{In \cite{Nijhoff} a discrete version of the Dubrovin equations was obtained associated with the finite-gap solutions of the lattice 
KdV system.},
\begin{equation}\label{eq:3.34}
\displaystyle\frac{1}{2\sqrt{R(\nu_{k}^{2})}}\displaystyle\frac{
\mathrm{d}\nu_{k}^{2}}{
\mathrm{d}t_{\lambda}}=-\displaystyle\frac{2\delta}{\alpha(\zeta)}\displaystyle\frac{
\mathfrak{n}(\zeta)}{
(\zeta-\nu_{k}^{2})\mathfrak{n}^{\prime}(\nu_{k}^{2})}.
\end{equation}
Hence by using the Lagrange interpolation formula for polynomials, we get
\begin{equation}\label{eq:3.35}
\sum\limits
_{k=1}^{g}\displaystyle\frac{
(\nu_{k}^{2})^{g-s}}{2\sqrt{R(\nu_{k}^{2})}}\displaystyle\frac{\mathrm{d}\nu_{k}^{2}}{
\mathrm{d}t_{\lambda}}=-\displaystyle\frac{2\delta}{\alpha(\zeta)}\zeta^{g-s},\quad(1\leq
s\leq g),
\end{equation}
which can be rewritten in the simple form
\begin{equation}\label{eq:3.36}
\displaystyle\frac{
\mathrm{d}\phi_{s}^{'}}{\mathrm{d}t_{\lambda}}=\{\phi_{s}^{'},\mathcal{F}_{\lambda}\}=-\displaystyle\frac{2\delta}{\alpha(\zeta)}\zeta^{g-s},\quad(1\leq
s\leq g)
\end{equation}
with the help of the quasi-Abel-Jacobi variables $\vec\phi^\prime=(\phi_{1}^{'},\ldots,\phi_{g}^{'})^{T}$,
\begin{equation}\label{eq:3.37}
\phi_{s}^{'}=\sum\limits
_{k=1}^{g}\displaystyle\int_{\mathfrak{p}_{0}}^{\mathfrak{p}(\nu_{k}^{2})}\displaystyle\frac{\zeta^{g-s}}{2\sqrt{R(\zeta)}}\mathrm{d}\zeta,\quad(1\leq
s\leq g),
\end{equation}
determined by the basis of holomorphic differentials in equation (\ref{eq:3.9}). Thus, we arrive at the following proposition:

\noindent \textbf{Proposition 3.3.} The invariants $E_{1},\ldots,E_{N}$ of the symplectic map $S_{\beta}$ are functionally independent throughout the phase space $\mathbb{R}^{2N}$.

\noindent \emph{Proof.}  Suppose $\Sigma_{j=1}^{N}c_{j}\mathrm{d}E_{j}=0$. Then
$\Sigma_{j=1}^{N}c_{j}\{\phi_{s}^{'},E_{j}\}=0,\forall s $. We shall now prove that $c_{j}=0,\forall j$. Substituting the expansion (\ref{eq:3.17}) into (\ref{eq:3.36}) we get
\begin{equation*}
\{\phi_{s}^{'},E_{j}\}=-\displaystyle\frac{2\delta}{\alpha^{\prime}(\alpha_{j}^{2})}(\alpha_{j}^{2})^{g-s},
\end{equation*}
by calculating the residues at points $\zeta=\alpha_{j}^{2} \ \ (1\leq j\leq N)$. Then the coefficient matrix $(\{\phi_{s}^{'},E_{j}\})_{N\times N}$ is non-degenerate since its determinant is Vandermonde determinant. This completes the proof. \hfill $\Box$

To summarise the results so far, the nonlinear map $S_{\beta}$ defined by (\ref{eq:3.12})
has been shown to be symplectic and integrable, posessing $N$ invarint functions $E_{1},\ldots, E_{N}$,
pairwise in involution and functionally independent on $\mathbb{R}^{2N}$.

\section{Evolution of the solutions to Q1 equation}\setcounter{equation}{0}

In the spirit of previous papers, \cite{Cao,Cao1,Cao2}, we will now discuss the actual
theta function solutions of the Q1 equation (\ref{eq:1.1}) arising from the integration of the symplectic map.
It is the compatibility of discrete flows that is essential for constructing these solutions, and we
will implement that on the relevant Baker-Akhiezer functions \cite{Baker,Akhiezer}. Thus, by considering the iteration of the
integrable symplectic map $S_{\beta}$ obtained in Section 3, we obtain a discrete phase flow $\big(p(m),q(m)
\big)=S^{m}_{\beta}(p_{0},q_{0})$, with $(p_0,q_0)\in \mathbb{R}^{2N}$ any initial value point. Here the number of iteration coincides with the lattice variable $m$. Then the
commutative relation (\ref{eq:3.3}) along the $S^{m}_{\beta}$-flow has the form,
\begin{equation}\label{eq:4.1}
\mathcal{L}_{m+1}(\lambda)\mathcal{D}^{(\beta)}_{m}(\lambda) =
\mathcal{D}^{(\beta)}_{m}(\lambda)\mathcal{L}_{m}(\lambda),
\end{equation}
where $\mathcal{L}_{m}(\lambda)=\mathcal{L}(\lambda;p(m),q(m))$, and $\mathcal{D}^{(\beta)}_{m}(\lambda)= \mathcal{D}^{(\beta)}(\lambda;b_{m})$ is the Darboux matrix given by (\ref{eq:2.1}) with discrete potentials $b_{m}, u_{m}$ satisfying
\begin{align}\label{eq:4.2}
\begin{split}
&\beta b_{m}=u_{m+1}-u_{m}, \ \ \mathrm{or}\\
&\beta b=\tilde{u}-u.
\end{split}
\end{align}
We note that both the eigenvalues $\pm \mathcal{H}_{\lambda}$ given by the formula (\ref{eq:3.5}) and the spectral curve $\mathcal {R}$ defined by equation (\ref{eq:3.8}) are invariant under the $S^{m}_{\beta}$-flow, since $\mathcal{F}_{\lambda}\big(p(m),q(m)\big)=\mathcal{F}_{\lambda}(p_{0},q_{0})$ by using equation (\ref{eq:3.16}). According to the Burchnall-Chaundy theory \cite{Burchnall,Burchnall1928,Naiman,Glazman}, we now investigate the common eigenvectors of the two matrix operators $\mathcal{L}_{m}(\lambda)$ and $\mathcal{D}^{(\beta)}_{m}(\lambda)$. Considering $\mathcal{D}^{(\beta)}_{m}(\lambda)$ as a shift operator, we suppose
\begin{equation}\label{eq:4.3}
h(m+1,\lambda) =
\mathcal{D}^{(\beta)}_{m}(\lambda)h(m,\lambda),
\end{equation}
where $h(m,\lambda)$ is the eigenvector corresponding to the eigenvalue $\lambda$. This is a linear equation, and then an alternative way of viewing $h(m,\lambda)$ is as a solution to the equation (\ref{eq:4.3}). Thus we discuss the fundamental solution matrix $ \mathcal{M}(m,\lambda)$ and find
\begin{equation}\label{eq:4.4}
 \mathcal{M}(m+1,\lambda) =
\mathcal{D}^{(\beta)}_{m}(\lambda) \mathcal{M}(m,\lambda), \ \
\mathcal{M}(0,\lambda) = I.
\end{equation}
Then by induction, the solution can be written as a matrix product chain
\begin{equation}\label{eq:4.5}
\mathcal{M}(m,\lambda) =
\mathcal{D}^{(\beta)}_{m-1}(\lambda)\mathcal{D}^{(\beta)}_{m-2}(\lambda)\ldots
\mathcal{D}^{(\beta)}_{0}(\lambda),
\end{equation}
which implies $\mathrm{det} \mathcal{M}(m,\lambda) = (\lambda^{2}-\beta^{2})^{m}$. Fortunately, the solution space of equation \eqref{eq:4.3} is invariant under the action of the algebra operator $\mathcal{L}_{m}(\lambda)$. In fact, by using the commutative relation \eqref{eq:4.1} we have
\begin{equation*}
(\mathcal{L}h)_{m+1}=\mathcal{L}_{m+1}(\mathcal{D}^{(\beta)}_{m}h_{m})=\mathcal{D}^{(\beta)}_{m}(\mathcal{L}h)_{m}.
\end{equation*}
Moreover, from equations \eqref{eq:4.1} and \eqref{eq:4.5} we obtain
\begin{equation}\label{eq:4.6}
\mathcal{L}_{m}(\lambda)\mathcal{M}(m,\lambda) =
\mathcal{M}(m,\lambda)\mathcal{L}_{0}(\lambda),
\end{equation}
which demonstrates that the evolution on the level of the Lax representation is nothing more than a
(matrix) similarity transformation.
Hence we consider the common eigenvectors $h_{\pm}(m,\lambda)$ associated with the eigenvalues $\lambda$ and $\pm \mathcal{H}_{\lambda}$, which satisfy the following formulas simultaneously:
\begin{align}\label{eq:4.7}
&\mathcal{L}_{m}(\lambda)h_{\pm}(m,\lambda)=\pm \mathcal{H}_{\lambda}h_{\pm}(m,\lambda),\\ \label{eq:4.8}
&h_{\pm}(m+1,\lambda) =
\mathcal{D}^{(\beta)}_{m}(\lambda)h_{\pm}(m,\lambda).
\end{align}
Since the rank of $\mathcal{L}_{m}(\lambda)\mp \mathcal{H}_{\lambda}I$ is 1, the eigenvector in each case is unique (up to constant factors). Thus, the simultaneous eigenvectors $h_{\pm}(m,\lambda)$ can be expressed as
\begin{align}\label{eq:4.9}
h_{\pm}(m,\lambda)={h_{\pm}^{(1)}(m,\lambda)\choose
h_{\pm}^{(2)}(m,\lambda)}=\mathcal{M}(m,\lambda){c_{\lambda}^{\pm}\choose
1}.
\end{align}
Substituting \eqref{eq:4.9} into \eqref{eq:4.7} and choosing $m=0$, we deduce
\begin{align}\label{eq:4.10}
c^{\pm}_{\lambda}=\frac{\mathcal{L}_0^{11}(\lambda)\pm
\mathcal{H}_{\lambda}}{\mathcal{L}_0^{21}(\lambda)}=\frac{-\mathcal{L}_0^{12}(\lambda)}{\mathcal{L}_0^{11}(\lambda)\mp
\mathcal{H}_{\lambda}}.
\end{align}
Referring to \cite{Krichever,Krichever1976,Krichever1977}, when the rank of the commuting pair, i.e., the dimension of the eigenspace of common eigenvectors, equals to 1, the associated equations of Lax type have finite-gap solutions. Here we shall investigate the common eigenvectors $h_{\pm}(m,\lambda)$ by using the Baker-Akhiezer functions, which can be expressed by theta functions on the hyperelliptic Riemann surface corresponding to the spectral curve $\mathcal{R}$ given by (\ref{eq:3.8}). Consequently, the discrete potentials $b_{m}, u_{m}$ in equation (\ref{eq:4.2}) can be reconstructed in terms of coefficients of the asymptotic expansions of these Baker-Akhiezer functions, which leads to the theta-function solutions for the Q1 equation (\ref{eq:1.1}) \cite{Cao,Cao1,Cao2}.

\noindent Technically, separating out the two cases: $m=2k-1,2k$, by using equation (\ref{eq:4.5}) and induction, we find that
 the following functions are polynomials of the argument $\zeta=\lambda^{2}$:
\begin{align*}
\mathcal{ M}^{21}(2k-1,\lambda), \lambda \mathcal{M}^{22}(2k-1,\lambda), \lambda \mathcal{M}^{21}(2k,\lambda), \mathcal{M}^{22}(2k,\lambda).
\end{align*}
Furthermore, it is easy to see that $\lambda c^{+}_{\lambda}$ and $\lambda c^{-}_{\lambda}$ are the values of a meromorphic function on $\mathcal{R}$,
\begin{equation*}
\mathcal{C}(\mathfrak{p})=\displaystyle\frac{\zeta<(\zeta-A^{2})^{-1}p_{0},q_{0}>+\delta\xi/\alpha(\zeta)}
{1+<(\zeta-A^{2})^{-1}Aq_{0},q_{0}>},
\end{equation*}
at the points $\mathfrak{p}(\lambda^{2})$ and $(\tau\mathfrak{p})(\lambda^{2})$, respectively. Thus, we can construct meromorphic functions $\mathfrak{h}^{(2)}(m,\mathfrak{p}), (m=2k-1,2k)$ on $\mathcal{R}$, i.e., Baker-Akhiezer functions, with the values at $\mathfrak{p}$ and $\tau\mathfrak{p}$ as
\begin{align}\label{eq:4.11}
\begin{split}
&\mathfrak{h}^{(2)}(2k-1,\mathfrak{p}(\lambda^{2}))=\lambda h_{+}^{(2)}(2k-1,\lambda), \ \
\mathfrak{h}^{(2)}(2k-1,\tau\mathfrak{p}(\lambda^{2}))=\lambda h_{-}^{(2)}(2k-1,\lambda),\\
&\mathfrak{h}^{(2)}(2k,\mathfrak{p}(\lambda^{2}))= h_{+}^{(2)}(2k,\lambda), \ \
\mathfrak{h}^{(2)}(2k,\tau\mathfrak{p}(\lambda^{2}))= h_{-}^{(2)}(2k,\lambda),
\end{split}
\end{align}
where
\begin{align}\label{eq:4.12}
\begin{split}
& \lambda h_{\pm}^{(2)}(2k-1,\lambda)=\mathcal{M}^{21}(2k-1,\lambda)\lambda c^{\pm}_{\lambda}+\lambda \mathcal{M}^{22}(2k-1,\lambda),\\
& h_{\pm}^{(2)}(2k,\lambda)=\lambda^{-1}\mathcal{M}^{21}(2k,\lambda)\lambda c^{\pm}_{\lambda}+ \mathcal{M}^{22}(2k,\lambda).
\end{split}
\end{align}
According to the theory of Riemann surface \cite{Farkas,Griffiths,Mumford}, we now find the zeros and poles for meromorphic functions $\mathfrak{h}^{(2)}(m,\mathfrak{p}), (m=2k-1,2k)$, which determine the expressions in terms of theta functions.

\noindent From equations \eqref{eq:4.9} and \eqref{eq:4.10}, we derive
\begin{equation}\label{eq:4.13}
h_{\pm}(m,\lambda)h_{\pm}^{T}(m,\lambda)=\displaystyle\frac{1}{\mathcal{L}^{21}_{0}(\lambda)}\mathcal{M}(m,\lambda)
[\mathcal{L}_{0}(\lambda) + \mathcal{H}_{\lambda}]\mathrm{i}\sigma_{2}\mathcal{M}^{T}(m,\lambda).
\end{equation}
Then by using equations \eqref{eq:3.32}, \eqref{eq:4.5} and \eqref{eq:4.6}, we find one entry of the above matrix equation \eqref{eq:4.13} satisfies
\begin{equation}\label{eq:4.14}
h^{(2)}_{+}(m,\lambda) \cdot h^{(2)}_{-}(m,\lambda)
=(\zeta-\beta^{2})^{m}\prod\limits_{j=1}^{N}
\displaystyle\frac{\zeta-\nu_{j}^{2}(m)}{\zeta-\nu_{j}^{2}(0)}.
\end{equation}
Thus, by equation \eqref{eq:4.11}, we have
\begin{align}\label{eq:4.15}
\begin{split}
&\mathfrak{h}^{(2)}(2k-1,\mathfrak{p}(\zeta))\mathfrak{h}^{(2)}(2k-1,\tau\mathfrak{p}(\zeta))=\zeta(\zeta-\beta^{2})^{2k-1}
\prod\limits_{j=1}^{N}\displaystyle\frac{\zeta-\nu_{j}^{2}(2k-1)}{\zeta-\nu_{j}^{2}(0)},\\
&\mathfrak{h}^{(2)}(2k,\mathfrak{p}(\zeta))\mathfrak{h}^{(2)}(2k,\tau\mathfrak{p}(\zeta))=(\zeta-\beta^{2})^{2k}
\prod\limits_{j=1}^{N}\displaystyle\frac{\zeta-\nu_{j}^{2}(2k)}{\zeta-\nu_{j}^{2}(0)},
\end{split}
\end{align}
which implies zeros and some poles, while the asymptotic behaviors in the vicinity of the infinity point on $\mathcal{R}$ will provide the remaining poles. Indeed, as $\lambda\sim \infty$, we obtain
\begin{equation}\label{eq:4.16}
c^{\pm}_{\lambda}=\pm\delta \lambda[1+O(\lambda^{-2})],
\end{equation}
by using equation \eqref{eq:4.10} and
\begin{align*}
&\mathcal{L}^{11}(\lambda)=O(\lambda^{-1}), \mathcal{L}^{12}(\lambda)=\delta^{2}\lambda^{2}[1+O(\lambda^{-4})],\\
&\mathcal{L}^{21}(\lambda)=1+O(\lambda^{-2}), \mathcal{H}_{\lambda}=\delta\lambda[1+O(\lambda^{-4})].
\end{align*}
Besides,
\begin{equation}\label{eq:4.17}
\mathcal{M}(m,\lambda)=\begin{pmatrix}\lambda^{m}Z_{m}^{11}[1+O(\lambda^{-2})] & \lambda^{m+1}Z_{m}^{12}[1+O(\lambda^{-2})] \\
 \lambda^{m-1}Z_{m}^{21}[1+O(\lambda^{-2})] & \lambda^{m}Z_{m}^{22}[1+O(\lambda^{-2})]\end{pmatrix},
\end{equation}
where
\begin{align}\label{eq:4.18}
&Z_{m}=\begin{pmatrix}Z_{m}^{11} & Z_{m}^{12} \\
 Z_{m}^{21}& Z_{m}^{22}\end{pmatrix}=\displaystyle\frac{1}{2\delta}\begin{pmatrix}\delta(z_{m}+z_{m}^{-1}) & \delta^{2}(z_{m}-z_{m}^{-1}) \\
z_{m}-z_{m}^{-1}&\delta(z_{m}+z_{m}^{-1})\end{pmatrix},\\ \label{eq:4.19}
&z_{m}=\Big(\displaystyle\frac{(b_{m-1}+\delta\beta)\cdots(b_{0}+\delta\beta)}{(b_{m-1}-\delta\beta)\cdots(b_{0}-\delta\beta)}\Big)^{1/2},
\end{align}
by equation \eqref{eq:4.5} and induction. Substituting \eqref{eq:4.16} and \eqref{eq:4.17} into \eqref{eq:4.12}, we obtain the asymptotic behaviour for $\mathfrak{h}^{(2)}(m,\mathfrak{p})$ near points $\infty_{+}$, $\infty_{-}$:
\begin{equation}\label{eq:4.20}
\begin{cases}
\mathfrak{h}^{(2)}(2k-1,\mathfrak{p})=z_{2k-1}\zeta^{k}[1+O(\zeta^{-1})],\ \ \mathfrak{p}\sim \infty_{+},\\
\mathfrak{h}^{(2)}(2k-1,\mathfrak{p})=z_{2k-1}^{-1}\zeta^{k}[1+O(\zeta^{-1})],\ \ \mathfrak{p}\sim \infty_{-}.
\end{cases}
 \end{equation}
\begin{equation}\label{eq:4.21}
\begin{cases}
\mathfrak{h}^{(2)}(2k,\mathfrak{p})=z_{2k}\zeta^{k}[1+O(\zeta^{-1})],\ \ \mathfrak{p}\sim \infty_{+},\\
\mathfrak{h}^{(2)}(2k,\mathfrak{p})=z_{2k}^{-1}\zeta^{k}[1+O(\zeta^{-1})],\ \ \mathfrak{p}\sim \infty_{-}.
\end{cases}
 \end{equation}

This leads to the following conclusion on the analytic behaviours of the Baker-Akhiezer functions.

\noindent
\textbf{Proposition 4.1.} The Baker-Akhiezer functions $\mathfrak{h}^{(2)}(2k-1,\mathfrak{p}),\mathfrak{h}^{(2)}(2k,\mathfrak{p})$ on $\mathcal{R}$ have the following divisors, respectively \cite{Griffiths,Mumford,Farkas}:
\begin{align}\label{eq:4.22}
\begin{split}
&\mathrm{Div}(\mathfrak{h}^{(2)}(2k-1,\mathfrak{p}))=\sum_{j=1}^{g}\big(\mathfrak{p}(\nu_{j}^{2}(2k-1))-\mathfrak{p}(\nu_{j}^{2}(0))\big)
+ \{\mathfrak{0}\}+(2k-1)\mathfrak{p}(\beta^{2})-k\infty_{+}-k\infty_{-},\\
&\mathrm{Div}(\mathfrak{h}^{(2)}(2k,\mathfrak{p}))=\sum_{j=1}^{g}\big(\mathfrak{p}(\nu_{j}^{2}(2k))-\mathfrak{p}(\nu_{j}^{2}(0))\big)
+2k\mathfrak{p}(\beta^{2})-k\infty_{+}-k\infty_{-}.
\end{split}
\end{align}
Let us now introduce the Abel-Jacobi variable $\vec\phi(m)=\mathscr
{A}(\Sigma_{k=1}^g\mathfrak p(\nu_k^{2}(m)))$ on $J(\mathcal R)=\mathbb C^g/\mathscr T$, with the help of the Able map $\mathscr
{A}$ given in \eqref{eq:3.11}. Resorting to the dipole technique developed by \cite{Toda}, equation \eqref{eq:4.22} implies
\begin{align*}
\begin{split}
&\vec{\phi}(2k-1) \equiv \vec{\phi}(0)+k\vec{\Omega}_{\beta}^{+}+(k-1)\vec{\Omega}_{\beta}^{-}+\vec{\Omega}_{0}^{-},\quad (\mathrm{mod}\mathscr {T}),\\
&\vec{\phi}(2k) \equiv \vec{\phi}(0)+k\vec{\Omega}_{\beta}^{+}+k\vec{\Omega}_{\beta}^{-},\quad (\mathrm{mod}\mathscr {T}),
\end{split}
\end{align*}
which can be rewritten as
\begin{equation}\label{eq:4.23}
\vec{\phi}(m) \equiv \vec{\phi}(0)+\displaystyle\frac{m+\Delta_{m}}{2}\vec{\Omega}_{\beta}^{+}+
\displaystyle\frac{m-\Delta_{m}}{2}\vec{\Omega}_{\beta}^{-}+\Delta_{m}\vec{\Omega}_{0}^{-},\quad (\mathrm{mod}\mathscr {T}),
\end{equation}
where $\vec{\Omega}_{\beta}^{+}=\int_{\mathfrak{p}(\beta^{2})}^{\infty_{+}}\vec{\omega}, \vec{\Omega}_{\beta}^{-}=\int_{\mathfrak{p}(\beta^{2})}^{\infty_{-}}\vec{\omega}$ and $\vec{\Omega}_{0}^{-}=\int_{\mathfrak{0}}^{\infty_{-}}\vec{\omega}$. Moreover, $\Delta_{j}$ is equal to 0 and 1 for even and odd $j$ respectively.

\noindent A usual argument leads to the expressions for the Baker-Akhiezer functions as \cite{Farkas,Griffiths,Mumford,Cao}
\begin{align}\label{eq:4.24}
&\mathfrak{h}^{(2)}(2k-1,\mathfrak{p})=C_{2k-1}\cdot\frac{\theta[-\mathscr{A}(\mathfrak{p})+\vec{\phi}(2k-1)+\vec{K}]}
{\theta[-\mathscr{A}(\mathfrak{p})+\vec{\phi}(0)+\vec{K}]}e^{\int_{\mathfrak{p}_0}^{\mathfrak{p}}
k\omega[\mathfrak{p}(\beta^{2}),\infty_{+}]+(k-1)\omega[\mathfrak{p}(\beta^{2}),\infty_{-}]+\omega[\mathfrak{0},\infty_{-}]},\\ \label{eq:4.25}
&\mathfrak{h}^{(2)}(2k,\mathfrak{p})=C_{2k}\cdot\frac{\theta[-\mathscr{A}(\mathfrak{p})+\vec{\phi}(2k)+\vec{K}]}
{\theta[-\mathscr{A}(\mathfrak{p})+\vec{\phi}(0)+\vec{K}]}e^{\int_{\mathfrak{p}_0}^{\mathfrak{p}}k\omega[\mathfrak{p}(\beta^{2}),\infty_{+}]
+k\omega[\mathfrak{p}(\beta^{2}),\infty_{-}]},
\end{align}
with $C_{2k-1},C_{2k}$ the constants; $\vec{K}$
the Riemann constant; $\omega[p,q]$ the dipole, a meromorphical differential having only
simple poles at $p,q$, with residues $+1,-1$, respectively.

\noindent From asymptotic behaviours \eqref{eq:4.20} and \eqref{eq:4.21}, we obtain
\begin{align}\label{eq:4.26}
&z_{2k-1}=C_{2k-1}\cdot\frac{\theta[-\mathscr{A}(\infty_{+})+\vec{\phi}(2k-1)+\vec{K}]}
{\theta[-\mathscr{A}(\infty_{+})+\vec{\phi}(0)+\vec{K}]}e^{\int_{\mathfrak{p}_0}^{\infty_{+}}
(k-1)\omega[\mathfrak{p}(\beta^{2}),\infty_{-}]+\omega[\mathfrak{0},\infty_{-}]}\cdot(r_{\beta}^{+})^{k},\\ \label{eq:4.27}
&z_{2k-1}^{-1}=C_{2k-1}\cdot\frac{\theta[-\mathscr{A}(\infty_{-})+\vec{\phi}(2k-1)+\vec{K}]}
{\theta[-\mathscr{A}(\infty_{-})+\vec{\phi}(0)+\vec{K}]}e^{\int_{\mathfrak{p}_0}^{\infty_{-}}k\omega[\mathfrak{p}(\beta^{2}),\infty_{+}]}
\cdot(r_{\beta}^{-})^{k-1}\cdot r_{0}^{-},\\ \label{eq:4.28}
&z_{2k}=C_{2k}\cdot\frac{\theta[-\mathscr{A}(\infty_{+})+\vec{\phi}(2k)+\vec{K}]}
{\theta[-\mathscr{A}(\infty_{+})+\vec{\phi}(0)+\vec{K}]}e^{\int_{\mathfrak{p}_0}^{\infty_{+}}
k\omega[\mathfrak{p}(\beta^{2}),\infty_{-}]}\cdot(r_{\beta}^{+})^{k},\\ \label{eq:4.29}
&z_{2k}^{-1}=C_{2k}\cdot\frac{\theta[-\mathscr{A}(\infty_{-})+\vec{\phi}(2k)+\vec{K}]}
{\theta[-\mathscr{A}(\infty_{-})+\vec{\phi}(0)+\vec{K}]}e^{\int_{\mathfrak{p}_0}^{\infty_{-}}k\omega[\mathfrak{p}(\beta^{2}),\infty_{+}]}
\cdot(r_{\beta}^{-})^{k},
\end{align}
where
\begin{eqnarray*}
r_{\beta}^{+}=\underset{\mathfrak{p} \rightarrow \infty^{+}}
{\mathrm{lim}}\displaystyle\frac{1}{\zeta(\mathfrak{p})}
e^{\int_{\mathfrak{p}_{0}}^{\mathfrak{p}}\omega[\mathfrak{p}(\beta^{2}),\infty_{+}]},\
\  r_{\beta}^{-}= \underset{\mathfrak{p} \rightarrow \infty^{-}}
{\mathrm{lim}}\displaystyle\frac{1}{\zeta(\mathfrak{p})}
e^{\int_{\mathfrak{p}_{0}}^{\mathfrak{p}}\omega[\mathfrak{p}(\beta^{2}),\infty_{-}]},\ \ r_{0}^{-}=\underset{\mathfrak{p} \rightarrow \infty^{-}}
{\mathrm{lim}}\displaystyle\frac{1}{\zeta(\mathfrak{p})}
e^{\int_{\mathfrak{p}_{0}}^{\mathfrak{p}}\omega[\mathfrak{0},\infty_{-}]}.
\end{eqnarray*}
Hence
\begin{align}\label{eq:4.30}
\begin{split}
z_{2k-1}^{2}=&\displaystyle\frac{
\theta[-\mathscr{A}(\infty_{+})+\vec{\phi}(2k-1)+\vec{K}]}
{\theta[-\mathscr{A}(\infty_{-})+\vec{\phi}(2k-1)+\vec{K}]} \cdot
\displaystyle\frac{\theta[-\mathscr{A}(\infty_{-})+\vec{\phi}(0)+\vec{K}]}{\theta[-\mathscr{A}(\infty_{+})+\vec{\phi}(0)+\vec{K}]}\cdot \\
&\cdot(\displaystyle\frac{r_{\beta}^{+}}{r_{\beta}^{-}})^{k-1}\cdot\displaystyle\frac{r_{\beta}^{+}}{r_{0}^{-}}
\cdot e^{\int_{\mathfrak{p}_{0}}^{\infty_{+}}(k-1)\omega[\mathfrak{p}(\beta^{2}),\infty_{-}]+\omega[\mathfrak{0},\infty_{-}]
-\int_{\mathfrak{p}_{0}}^{\infty_{-}}k\omega[\mathfrak{p}(\beta^{2}),\infty_{+}]}
, \\
z_{2k}^{2}=&\displaystyle\frac{
\theta[-\mathscr{A}(\infty_{+})+\vec{\phi}(2k)+\vec{K}]}
{\theta[-\mathscr{A}(\infty_{-})+\vec{\phi}(2k)+\vec{K}]} \cdot
\displaystyle\frac{\theta[-\mathscr{A}(\infty_{-})+\vec{\phi}(0)+\vec{K}]}{\theta[-\mathscr{A}(\infty_{+})+\vec{\phi}(0)+\vec{K}]}\cdot \\
&\cdot(\displaystyle\frac{r_{\beta}^{+}}{r_{\beta}^{-}})^{k}
\cdot e^{\int_{\mathfrak{p}_{0}}^{\infty_{+}}k\omega[\mathfrak{p}(\beta^{2}),\infty_{-}]
-\int_{\mathfrak{p}_{0}}^{\infty_{-}}k\omega[\mathfrak{p}(\beta^{2}),\infty_{+}]}.
\end{split}
\end{align}
Then by using equation \eqref{eq:4.19}, we have
\begin{align}\label{eq:4.31}
\begin{split}
\displaystyle\frac{b_{2k-1}+\delta\beta}{b_{2k-1}-\delta\beta}=\displaystyle\frac{z_{2k}^{2}}{z_{2k-1}^{2}}=&\displaystyle\frac{
\theta[-\mathscr{A}(\infty_{+})+\vec{\phi}(2k)+\vec{K}]}
{\theta[-\mathscr{A}(\infty_{-})+\vec{\phi}(2k)+\vec{K}]} \cdot
\displaystyle\frac{\theta[-\mathscr{A}(\infty_{-})+\vec{\phi}(2k-1)+\vec{K}]}{\theta[-\mathscr{A}(\infty_{+})+\vec{\phi}(2k-1)+\vec{K}]}\cdot \\
&\cdot\displaystyle\frac{r_{0}^{-}}{r_{\beta}^{-}}
\cdot e^{\int_{\mathfrak{p}_{0}}^{\infty_{+}}\omega[\mathfrak{p}(\beta^{2}),\infty_{-}]-\omega[\mathfrak{0},\infty_{-}]}, \\
\displaystyle\frac{b_{2k}+\delta\beta}{b_{2k}-\delta\beta}=\displaystyle\frac{z_{2k+1}^{2}}{z_{2k}^{2}}=&\displaystyle\frac{
\theta[-\mathscr{A}(\infty_{+})+\vec{\phi}(2k+1)+\vec{K}]}
{\theta[-\mathscr{A}(\infty_{-})+\vec{\phi}(2k+1)+\vec{K}]} \cdot
\displaystyle\frac{\theta[-\mathscr{A}(\infty_{-})+\vec{\phi}(2k)+\vec{K}]}{\theta[-\mathscr{A}(\infty_{+})+\vec{\phi}(2k)+\vec{K}]}\cdot \\
&\cdot\displaystyle\frac{r_{\beta}^{+}}{r_{0}^{-}}
\cdot e^{\int_{\mathfrak{p}_{0}}^{\infty_{+}}\omega[\mathfrak{0},\infty_{-}]-\int_{\mathfrak{p}_{0}}^{\infty_{-}}\omega[\mathfrak{p}(\beta^{2}),\infty_{+}]},
\end{split}
\end{align}
which can be put in a unified form by introducing the function
\begin{align}\label{eq:4.32}
\begin{split}
\Upsilon_{m}=\displaystyle\frac{b_{m}+\delta\beta}{b_{m}-\delta\beta}=&\displaystyle\frac{
\theta[-\mathscr{A}(\infty_{+})+\vec{\phi}(m+1)+\vec{K}]}
{\theta[-\mathscr{A}(\infty_{-})+\vec{\phi}(m+1)+\vec{K}]} \cdot
\displaystyle\frac{\theta[-\mathscr{A}(\infty_{-})+\vec{\phi}(m)+\vec{K}]}{\theta[-\mathscr{A}(\infty_{+})+\vec{\phi}(m)+\vec{K}]}\cdot \\
&\cdot\displaystyle\frac{(r_{\beta}^{+})^{\Delta_{m+1}}}{(r_{\beta}^{-})^{\Delta_{m}}}\cdot(r_{0}^{-})^{(-1)^{\Delta_{m+1}}}\cdot \\
&\cdot e^{\int_{\mathfrak{p}_{0}}^{\infty_{+}}\Delta_{m}\omega[\mathfrak{p}(\beta^{2}),\infty_{-}]+(-1)^{\Delta_{m}}\omega[\mathfrak{0},\infty_{-}]
-\Delta_{m+1}\int_{\mathfrak{p}_{0}}^{\infty_{-}}\omega[\mathfrak{p}(\beta^{2}),\infty_{+}]},
\end{split}
\end{align}
from which we have by inverting the definition
\begin{equation}\label{eq:4.33}
b_{m}=\displaystyle\frac{\delta\beta(\Upsilon_{m}+1)}{\Upsilon_{m}-1}.
\end{equation}
Substituting equation \eqref{eq:4.2} into \eqref{eq:4.33} we arrive at a recursive relation for the potential $u_{m}$, in terms of theta functions,
\begin{equation}\label{eq:4.34}
u_{m+1}-u_{m}=\displaystyle\frac{\delta\beta^{2}(\Upsilon_{m}+1)}{\Upsilon_{m}-1}.
\end{equation}

In order to get the algebro-geometric solutions to equation \eqref{eq:1.1}, we introduce two distinct and non-zero lattice parameters $\beta_{1},\beta_{2}$.
According to the results in Section 3, we obtain two commuting integrable maps $S_{\beta_1},\,S_{\beta_2}$ since they share the same invariants $E_{1},\ldots,E_{N}$ \cite{Veselov,Veselov1,Bruschi,Suris,Cao2}. Their iteration give rise to commuting discrete flows $S_{\beta_{1}}^{m}$ and $S_{\beta_{2}}^{n}$.
Consequently the following function is well-defined on the $\mathbb{Z}^2$ lattice:
\begin{align}\label{eq:4.35}
\begin{split}
\big(p(m,n),q(m,n)\big)&=S_{\beta_{1}}^{m}S_{\beta_{2}}^{n}(p_{0},q_{0})=S_{\beta_{1}}^{m}\big(p(0,n),q(0,n)\big)\\
&=S_{\beta_{2}}^{n}S_{\beta_{1}}^{m}(p_{0},q_{0})=S_{\beta_{2}}^{n}\big(p(m,0),q(m,0)\big).
\end{split}
\end{align}
Furthermore, by equations \eqref{eq:3.1} in the two special cases and \eqref{eq:4.2}, the $j$-th component satisfies
\begin{align}\label{eq:4.36}
&\begin{pmatrix}\tilde{p}_{j}\\\tilde{q}_{j}\end{pmatrix}
=(\alpha_{j}^{2}-\beta_{1}^{2})^{-1/2}\mathcal{D}^{(\beta_{1})}(\alpha_{j};b_{1})\begin{pmatrix}p_{j}\\ q_{j}\end{pmatrix},\ \
b_{1}=\displaystyle\frac{\tilde{u}-u}{\beta_{1}}  \\ \label{eq:4.37}
&\begin{pmatrix}\bar{p}_{j}\\ \bar{q}_{j}\end{pmatrix}
=(\alpha_{j}^{2}-\beta_{2}^{2})^{-1/2}\mathcal{D}^{(\beta_{2})}(\alpha_{j};b_{2})\begin{pmatrix}p_{j}\\
q_{j}\end{pmatrix}, \ \ b_{2}=\displaystyle\frac{\bar{u}-u}{\beta_{2}},
\end{align}
which are compatible on account of the commutative relation between the maps $S_{\beta_1}$ and $S_{\beta_2}$. Thus $\bar{D}^{(\beta_{1})}D^{(\beta_{2})}=\tilde{D}^{(\beta_{2})}D^{(\beta_{1})}$. Then by using equation (\ref{eq:2.3}), the evolution of the recursive relation (\ref{eq:4.34}) along the flows $S_{\beta_{1}}^{m}$ and $S_{\beta_{2}}^{n}$ leads to the solutions for Q1 lattice equation (\ref{eq:1.1}) in the form
as expressed in the following.

\noindent \textbf{Proposition 4.2.} The Q1 equation (\ref{eq:1.1}) has special solutions $u=u_{m,n}$ satisfying
\begin{equation}\label{eq:4.38}
u_{m+1,n}-u_{m,n}=\displaystyle\frac{\delta\beta_{1}^{2}(\Upsilon_{m,n}+1)}{\Upsilon_{m,n}-1},
\end{equation}
where
\begin{align}\label{eq:4.39}
\begin{split}
\Upsilon_{m,n}=&\displaystyle\frac{
\theta[-\mathscr{A}(\infty_{+})+\vec{\phi}(m+1,n)+\vec{K}]}
{\theta[-\mathscr{A}(\infty_{-})+\vec{\phi}(m+1,n)+\vec{K}]} \cdot
\displaystyle\frac{\theta[-\mathscr{A}(\infty_{-})+\vec{\phi}(m,n)+\vec{K}]}{\theta[-\mathscr{A}(\infty_{+})+\vec{\phi}(m,n)+\vec{K}]}\cdot \\
&\cdot\displaystyle\frac{(r_{\beta_{1}}^{+})^{\Delta_{m+1}}}{(r_{\beta_{1}}^{-})^{\Delta_{m}}}\cdot(r_{0}^{-})^{(-1)^{\Delta_{m+1}}}\cdot \\
&\cdot e^{\int_{\mathfrak{p}_{0}}^{\infty_{+}}\Delta_{m}\omega[\mathfrak{p}(\beta_{1}^{2}),\infty_{-}]+(-1)^{\Delta_{m}}\omega[\mathfrak{0},\infty_{-}]
-\Delta_{m+1}\int_{\mathfrak{p}_{0}}^{\infty_{-}}\omega[\mathfrak{p}(\beta_{1}^{2}),\infty_{+}]},\\
\vec{\phi}(m,n) \equiv &\vec{\phi}(0,0)+\displaystyle\frac{m+\Delta_{m}}{2}\vec{\Omega}_{\beta_{1}}^{+}+
\displaystyle\frac{m-\Delta_{m}}{2}\vec{\Omega}_{\beta_{1}}^{-}+\\
&\displaystyle\frac{n+\Delta_{n}}{2}\vec{\Omega}_{\beta_{2}}^{+}+
\displaystyle\frac{n-\Delta_{n}}{2}\vec{\Omega}_{\beta_{2}}^{-}+
(\Delta_{m}+\Delta_{n})\vec{\Omega}_{0}^{-},\quad (\mathrm{mod}\mathscr {T}),
\end{split}
\end{align}
and $\vec{\Omega}_{\beta_{j}}^{+}=\int_{\mathfrak{p}(\beta_{j}^{2})}^{\infty_{+}}\vec{\omega}, \vec{\Omega}_{\beta_{j}}^{-}=\int_{\mathfrak{p}(\beta_{j}^{2})}^{\infty_{-}}\vec{\omega},\ \ j=1,2.$

\noindent Besides,
\begin{equation}\label{eq:4.40}
u_{m,n+1}-u_{m,n}=\displaystyle\frac{\delta\beta_{2}^{2}(\Theta_{m,n}+1)}{\Theta_{m,n}-1},
\end{equation}
where
\begin{align}\label{eq:4.41}
\begin{split}
\Theta_{m,n}=&\displaystyle\frac{
\theta[-\mathscr{A}(\infty_{+})+\vec{\phi}(m,n+1)+\vec{K}]}
{\theta[-\mathscr{A}(\infty_{-})+\vec{\phi}(m,n+1)+\vec{K}]} \cdot
\displaystyle\frac{\theta[-\mathscr{A}(\infty_{-})+\vec{\phi}(m,n)+\vec{K}]}{\theta[-\mathscr{A}(\infty_{+})+\vec{\phi}(m,n)+\vec{K}]}\cdot \\
&\cdot\displaystyle\frac{(r_{\beta_{2}}^{+})^{\Delta_{n+1}}}{(r_{\beta_{2}}^{-})^{\Delta_{n}}}\cdot(r_{0}^{-})^{(-1)^{\Delta_{n+1}}}\cdot \\
&\cdot e^{\int_{\mathfrak{p}_{0}}^{\infty_{+}}\Delta_{n}\omega[\mathfrak{p}(\beta_{2}^{2}),\infty_{-}]+(-1)^{\Delta_{n}}\omega[\mathfrak{0},\infty_{-}]
-\Delta_{n+1}\int_{\mathfrak{p}_{0}}^{\infty_{-}}\omega[\mathfrak{p}(\beta_{2}^{2}),\infty_{+}]}.
\end{split}
\end{align}

This proposition forms the main and final result of the paper. We remark that based on \eqref{eq:4.38} and \eqref{eq:4.40} the solutions to Q1 equation (\ref{eq:1.1}), in terms of theta functions, should be integrated in order to find $u=u_{m,n}$ in the form
\begin{align}\label{eq:4.42}
\begin{split}
u&=u_{0,n}+\sum_{j=1}^{m}\big(u_{j,n}-u_{j-1,n}\big)\\
&=u_{m,0}+\sum_{j=1}^{n}\big(u_{m,j}-u_{m,j-1}\big).
\end{split}
\end{align}
However, it is not clear yet that the sum (\ref{eq:4.42}) can be explicitly computed in closed form.

\section{Conclusion}\setcounter{equation}{0}

In this paper we constructed algebro-geometric solutions of the Q1 equation \eqref{eq:1.1} in a novel way different from earlier approaches
applied in e.g. KdV type systems, \cite{Cao,Cao1,Cao2} where the continuous spectral problems associated with the integrable Hamiltonian systems in the
Liouville sense are essential. Here only the lattice equation is the starting point which gives rise to the purely discrete Lax pair by means of
the multidimensional consistency. From the latter we deduce the compatibility relations, as well as the relevant spectral curve and the
asscoaited hyperelliptic Riemann surface. Moreover, the integrability for the symplectic maps are studied with the help of the $r$-matrix.
An outstanding new feature in the present approach is the revelation that the discrete systems themselves without the continuous integrability
provides enough information for calculating the exact analytic solutions by the finite-gap technique.
Thus, in a sense it is justified to consider the discrete integrability as the more fundamental aspect, which can subsequently be used to investigate the
associated continuous integrable systems.

We note, however, that the solutions are obtained in a `derived' form, which to obtain the solution, should still be integrated (in the discrete sense), to get
the algebro-geometric solutions for Q1 lattice equation in explicit form. Whether or not those forms can be explicitly integrated remains an open problem,
but there are precedents in the case of soliton solutions where that can be done (see \cite{AtkHietNij}). However, the latter, which has never been
achieved in the finite-gap case, is a matter for future investigation beyond the scope of the present paper.

We finish by expressing our confidence that the techniques used in the present paper for the Q1 equation can be readily extended to the remaining
equations in the ABS list which are beyond the KdV type class, and notably to the problem of constructing algebro-geometric solutions of the Q4
equation which figures at the top of the list.

\vspace*{0.5cm} \noindent{\bf Acknowledgments}

This work is supported by National Natural Science Foundation of China (Grant
Nos. 11426206; 11501521), State Scholarship Found of China (CSC No. 201907045035), and Graduate Student Education Research Foundation
of Zhengzhou University (Grant No. YJSXWKC201913). We would like to express many thanks to Prof. Da-jun Zhang for helpful discussions.

\end{document}